\documentstyle[prl,aps]{revtex}

\begin{document}
\title{Why does a protein fold?}
\author{R. A Broglia$^{1,2,3}$ and G. Tiana$^{3}$}
\address{$^1$ Department of Physics, University of Milano, via
Celoria 16, I-20133 Milano, Italy}
\address{$^2$ Istituto Nazionale di Fisica Nucleare, via
Celoria 16, I-20133 Milano, Italy}
\address{$^3$ The Niels Bohr Institute, University of Copenhagen, 
Bledgamsvej 17, DK-2100 Copenhagen, Denmark}
\date{\today}
\maketitle

\bigskip

\begin{abstract}
With the help of lattice Monte Carlo modelling of heteropolymers, we show
that the necessary condition for a
protein to fold on short call is to proceed through partially folded
intermediates. These elementary structures are formed at an early stage in the
folding process and contain, at the local level, essentially all of the amino
acids found in the folding core (transition state) of the protein, providing
the local guidance for its formation. 
The sufficient condition for the protein 
to fold is that the designed sequence has
an energy, in the native conformation, below $E_c$ (the lowest energy of the
structurally dissimilar compact conformations) where it has not to compete with
the bulk of misfolded conformations. Sequences with energy close to $E_c$
can display prion--like behaviour, folding to two structurally dissimilar
conformations, one of them being the native.
\end{abstract}
\bigskip

We wish to suggest a novel model for protein folding, where  
the building blocks which control the dynamics of the designed sequences
are partially folded intermediates. 
Starting form a random coil (Fig. 1(a) ), they  are formed only after
$\approx 10^2$ steps of Monte Carlo (MC) simulations (Fig. 1(b) ), when some
of the most strongly interacting amino acids establish their local contacts.
They achieve $\approx 90-95\%$ stability after $10^5$ MC steps, and when they 
assemble together after $\approx 10^6$ MC steps (Fig. 1(c) ) they form  
the (post--critical) folding core \cite{p4,nucleus} of the notional protein, from   
which it reaches the native conformation (Fig. 1(d) ) in a short time  
($\approx 10^3$ MC steps), provided the energy of the system is lower 
than $E_c$. 
Partially folded intermediates and not the individual
monomers thus take care, through local guidance and non--local long range
correlations (bonding between partially folded intermediates),
of the process of protein folding, as testified by the disruptive effect
mutations which affect the stability of these structures have on the
folding ability of the designed sequence 
\cite{mut,aggreg,n1,n2,n3,n4} (cf. also \cite{note1}). 

The fast formation of few partially folded intermediates, and of their
bonding, reduces, in a conspicuous way, the number
of conformations that need to be searched (in case of the chain considered
in Fig. 1 to $10^{12}$ as compared to
$10^{24}$ for the random--coil), leading to the resolution of the Levinthal
paradox. It is also a very efficient way to squeeze entropy from the
system ($\approx50\%$) at the very early stages of the folding process, and
to repeat this feat when the partially folded intermediates come together
to form the folding nucleus (cf also ref. \cite{aggreg}), at which stage the
integrated decrease of entropy amounts to a large (in the case
of Fig. 1, of the order of $80\%$) fraction of the
original random--coil value.

The numbers quoted in the first paragraph were obtained using a lattice model 
of proteins studied earlier by us \cite{mut,aggreg,prl} and others
\cite{s2,klimov2,socci}. The model sequences are composed of amino
acids of 20 types and containing 36 monomers, 
which interact through contact energies
obtained from a statistical analysis of real proteins \cite{mj}, the associated
standard deviation of the interaction energies between different amino acid
types being $\sigma=0.3$. 
From very long Monte Carlo runs
($\approx 10^9$ MC steps) a sequence has been found with a sufficiently 
low energy in the native conformation \cite{note2},
which in the units we use ($RT_{room}=0.6\; kcal/mol$) is equal to
$E_n=-17.13$, to be compared to the lowest energy of the structurally
dissimilar part of the spectrum $E_c=-14$ (with standard deviation
$\sigma_c=0.2$), obtained through a set of low temperature MC samplings in
conformational space. 
While all sequences lying below $E_c$ (of the order of $10^{10}$
\cite{prl})
eventually fold, in keeping with the fact that they share a (small) number
of conserved contacts (folding nucleus), the folding time
is correlated with the corresponding energy gap
(Table 1). 
To state that the ability a notional protein has to fold, is connected with
the presence of a small number of conserved contacts or, equivalently, of conserved
amino acids \cite{mirnyy},
is tantamount to saying that foldability is connected with the presence of
a small number of partially folded intermediates. 
In fact, although most of the conserved contacts found in the folding nucleus
of ref. \cite{mirnyy} are non--local, few of them are local.
These few contacts stabilize the partially folded intermediates already at
the initial stage of the folding process.
It is then natural that the non--local contacts of the folding
nucleus arize from the assembling together of the partially folded
intermediates.
Because these local structures are both few and strongly
interacting as they are mediated, by the few, strongly interacting, 
amino acids occupying "hot" sites in the protein (cf. caption to Fig. 1
and \cite{mut}), they can come together  both fast ($\approx 10^5-10^6$MC steps)
and in a unique fashion, to form the (post--critical) 
folding core of the protein.
Consequently, the findings displayed in Fig. 1 agree in
detail with the result of ref. \cite{mirnyy} providing it 
a simple microscopic picture.

In spite of the difference in language, it also agrees with the findings of ref.
\cite{thirumalai}. In fact, while the stability of the folding intermediates
is not $100\%$, the corresponding contacts (cf. Fig. 2) 
are operative with an incidence
which is much higher than that associated with the non--conserved contacts
(cf. also Fig. 5 of ref. \cite{thirumalai}). 
In keeping with the definition of the transition states (which, in
the present case, are $\approx 10^4$) as those
in which the protein has equal probability to proceed to the native
conformation as it has to unfold, not all the conserved contacts are, in these
states, operative with probability 1. In this sense, good folders can
fold in different manners (different transition states) \cite{thirumalai}.
On the other hand, any good
folder passes, with probability 1, through the (post--critical) 
folding core conformation (Fig. 1(c) ){\it
en route} to the native structure.

In order to investigate the dynamical behaviour of sequences with energy
$E_n\leq E\lesssim E_c$ (that is sequences which can also be marginally stable), a
database of (composition conserving) sequences of specified energy has been created making use of a
Monte Carlo algorithm. The database is divided in 6 groups whose elements have
energy  $-17.00<E<-16.50$, $-16.50<E<-15.00$, ... , $-14.50<E<-14.00$, each
group containing 500 sequences. For each group the Monte Carlo selection has
been performed at a temperature ($T=0.28$) such that the average energy lies in the
associated energy interval. 

Essentially
all sequences ($92\%$) with $E<E_c$ fold in rather homogeneous times
(cf. Table
1), a time which is much shorter than that associated with a random search
in the space of compact conformations ($\sim 10^{12}$), let alone the full
space of conformations ($\sim 10^{24}$). The sequences fold either 
to the native structure or to a
unique structure with a value of the similarity parameter 
(defined as the ratio of native contacts of
a given conformation to the total number of contacts) $q>0.6$. This process
takes always place
through partially folded intermediates, a result which seems to find strong
experimental support (cf. e.g. \cite{baldwin} and refs. therein). 
To be noted that for
a given native structure,  all designed
sequences are characterized by a very limited choice of partially 
folded intermediates \cite{isles}. For example, in the case of
the native conformation chosen for the analysis (Fig. 1(d) ), 
these local substructures 
involve monomers 3--6, 11--14 and 27--30 \cite{aggreg} (the only
other choice for partially folded intermediates involves monomers
 2--7, 11--14, 16--21)\cite{note3}. We find that in 
the folding process, all sequences with  energy $E\lesssim E'_c$ where 
$E'_c=E_c-n\sigma$ ($n=1-2$), undergo a first order--like transition
\cite{privalov},
the transition state being characterized by values of the
conformational order parameter q  which range, depending on  
the sequence, between $0.45$ and $0.70$.
As expected from the definition of $E_c$, 
for sequences with energy close to $E_c$, the native conformation
starts competing with other conformations.
In fact, 
sequences which do not fold ($8\%$ of the total database) are concentrated
in the energy range  $-14.50<E<-14.00$. Of them, $2.7\%$ behave like random
sequences, while $5.3\%$ display an unexpected prion--like behaviour, folding
either to the native state or to another unique conformation.
The fact that also these sequences display
(according to simulations performed in the range of $10^8$ MC steps) a
first order--like transition, suggests that a mechanism of kinetic partitioning
is active, in the sense that, in the folding time scale, the unfolded and
only one of the two possible folded conformations
play a role in each simulation. In other words, prion--like sequences
behave as if, at the very early stages of the folding process, one of the two
possible folded conformations was selected (cf. also ref. \cite{prion}). In keeping with these results,
and to the extent that lattice simulations do describe "wild type" proteins,
one could argue that the mere existence of prions testifies to the central role
partially folded intermediates (only "intelligent" structures operative at
the very early stage of the compaction process) play in the folding of proteins.

We have also found that many sequences with $E\gtrsim E_c$ 
can still fold to the native conformation,
while a consistent part of them again show prion--like behaviour. For example, among sequences
with $-14.00<E<-13.50$, $53\%$ of them fold to the native state, $8\%$ fold to a unique
conformation, similar to the native state ($q>0.6$), $28\%$ display prion--like
behaviour,
while $11\%$ do not fold.  Among sequences with 
energy lying in the interval $-13.50<E<13.00$, $40\%$ can still fold to the native
state within $2\cdot 10^8$ MC steps, while sequences which reach a unique
conformation, different from the native, 
drop to $3\%$ and those displaying prion--like behaviour become $22\%$.

Partially folded intermediates are also found to be present in the compaction of
sequences displaying an
energy, in the native conformation, much larger than $E_c$.
Making again use of a Monte Carlo algorithm, we have investigated the average
native energy of sequences at different selective temperatures $T_s$, together
with the average energy of the partially folded intermediates and of the
folding nucleus. The results are
displayed in Fig. 3. Both the partially folded intermediates and the folding core undergo something
resembling a first order phase transition (strongly blurred by fluctuations, in keeping 
with the fact that the system is small) at $T_s\approx 0.3$,
while, in the same range of selective 
temperatures, the overall sequence undergoes a second order transition.
The energy $E_c$ corresponds to a selective temperature $T_s=0.09$, which is
far below the phase transition temperature. Consequently, partially folded intermediates
and the folding core are, in average, present in all the sequences with energy
as high as $E=-9$. In spite of the fact that some of these sequences 
are able to fold, folding events are rare at these energies, in keeping with
the fact that the folded state is
immersed in a dense background of states associated with random sequences and
thus of misfolded conformation.

The presence
of a specific set of very favorable interactions, the folding core \cite{p4,nucleus}, 
which is built out of
the partially folded intermediates and which depends on
the geometry of the target structure, and consequently is missed by the mean
field description, can lower the energy of the native state below $E_c$. 
In other words, the ground state energy of a sequence 
can be written as $E=E_{core}+E_{oth}$, where $E_{core}$ is the energy
of the folding nucleus, an energy which, according to a first order transition
interpretation of Fig. 3 can be either
$0$ or $J$ (with $J=-7$). The energy $E_{oth}$ of the non--core residues are distributed
according to the Random Energy Model \cite{rem}, the lowest of them being
$E''_c$ (which is higher than $E_c$ because it contains less residues). Setting 
$E''_c=E_c (n-n_{core})/n$ with $n=40$ the total number of contacts, and
with $n_{core}=9$ the contacts belonging to the folding core (cf. Fig. 1(c) ), one obtains
$E''_c=-10.9$. Then, the energy of the lowest sequence
should be $E_{opt}=-17.9$ (to be compared to the value $-17.13$ we obtained 
in MC simulations), 
so that the gap of the best sequence is $\delta_{opt}=E_c-E_{opt}=3.9$ 
(to be compared to
$\delta_{opt}=3.13$, the outcome of MC simulations). In keeping with this discussion, we find that 
there are
two sets of sequences. One, which in the compaction process
does not display  partially folded intermediates and thus a folding
core, spanning the energy interval $E''_c<E<0$. Another, spanning the energy
interval $E''_c+J<E<J$, which in the compaction
process form partially folded intermediates.
Sequences of this type with energies $E\lesssim E_c$ fold in times which are,
within an order of magnitude, essentially the same (Table 1). 
Within the present model, this is a rather natural result 
due to the fact that the folding time
is, to a large extent, determined by the time it takes 
for the partially folded
intermediates to assemble together into the folding core, 
and to the result that the partially
folded intermediates are the same for all sequences with $E<E_c$.

To the question: why does a protein fold?, the answer seems to be: 
because it proceeds through early formed
local structures, partially folded intermediates 
(efficient way to squeeze entropy from the chain)
carriers of the information concerning the folding core they
form by assembling
together, thus lowering
the energy of the system below the threshold energy of random sequences, 
where the system has
not to compete with the bulk of misfolded conformation.

\begin{table}
\begin{tabular}{|c|c|c|c|}
\hline
E interval  & $<t>$         & $\sigma_t$ & $\%$ native \\\hline
$-17/-16.5$ & $5\cdot 10^5$ & $2\cdot 10^5$ & 100\\
$-16.5/-16$ & $6\cdot 10^5$ & $2\cdot 10^5$ & 100\\
$-16/-15.5$ & $2\cdot 10^6$ & $2\cdot 10^6$ & 98\\
$-15.5/-15$ & $2\cdot 10^6$ & $2\cdot 10^6$ & 81\\
$-15/-14.5$ & $3\cdot 10^6$ & $5\cdot 10^6$ & 75\\
$-14.5/-14$ & $6\cdot 10^6$ & $9\cdot 10^6$ & 61 

\end{tabular}
\caption{For sequences belonging to the intervals of energy indicated in the
first column, it is listed the average first passage time needed for sequences to reach the ground state
conformation ($q>0.6$), the associated standard deviation, and the
percentage of sequences folding to the exact target structure ($q=1$).}
\end{table}

\begin{figure}
\caption{Snapshots of the folding of the sequence $S'_{36}$ (cf. footnote number
2), whose energy in the
native conformation is $E_n=-17.13$. Starting from a random conformation (a),
the system forms after $\approx 10^2$ MC steps partially folded intermediates
(b), involving three sets of four amino acids (3--6, 11--14, 17--30), whose
stability is provided by the bonding indicated by dotted lines. When the
partially folded intermediates come together to form the folding core (indicated
by dotted and dashed lines) after
$7\cdot 10^5$ MC steps (c), the system has reached the transition state from
which it folds to the native conformation after only $10^3$ MC steps (d).
The amino acids participating in the bonding of the partially folded
intermediates (dotted lines) are among some of the most strongly interacting
amino acids, which occupy, in the native conformation (d), "hot" and "warm"
sites [3] indicated by dark-- and light--gray beads, respectively. The monomers number 1 and
number 36 of the sequence $S'_{36}$ are indicated for each conformation.}
\end{figure}

\begin{figure}
\caption{
Dynamics of contact formation for two MC simulations of the folding of the 
S$_{36}$ sequence \protect\cite{note2}.
With a, b, c we have labeled the contact $27-30$, $11-14$ and $3-6$,
stabilizing the partially folded intermediates (cf. Fig. 1). With a solid
dot along the vertical axis we label (from top to bottom) the contacts:
5--28, 3--30, 14--27, 6--11, 13--28, 6--27, 12--5, 4--29 forming the
folding core. In the simulation associated with the results displayed in the
left panel, the protein folds in $5.1\cdot 10^5$ MC steps, while in that
associated with the right pannel it folds in $6.5\cdot 10^5$ MC steps.}
\end{figure}

\begin{figure}
\caption{The average energy of sequences in the native conformation is shown
(solid curve) as a function of the selective temperature.
The contact energy stored within the partially folded
intermediates is displayed in terms of a dotted curve, while the energy
arising from the interaction between the three partially folded intermediates
is shown in terms of a dashed curve. 
}
\end{figure}


\begin{thebibliography}{99}
\bibitem{note1} {Within the present 
scenario of protein folding the question of the relative
importance of local-- versus non--local contacts (cf. e.g.
\cite{gov,chan,plaxco} 
and refs. therein) is somewhat trascended. In fact, they   
become complementary expressions of the same reality.}
\bibitem{note2} {This sequence is
$S'_{36}$=YPDLT\-KWHA\-MEAGK\-IRFSV\-PDACLN\-GEGIR\-QVTLSN.
In previous papers  the sequence used as optimal was $S_{36}$ (cf. e.g. \cite{mut,aggreg,s2,prl}), its energy in
the native conformation being $-16.5$. To be noted that the behaviour of all
sequences with $E\ll E_c$ is essentially identical.}
\bibitem{note3} {Partially 
folded intermediates involve very few monomers. Consequently it is
difficult to associate them to macroscopic measurable quantities and, in
general, to detect them directly from experiment,
except if their formation is associated with a local minimum of
the overall energy (for example "three--states folding"). 
On the other hand, the
monomers participating in the formation of partially folded intermediates are
among the strongest interacting amino acids. Consequently, a possible way to
learn about partially folded intermediates, is to perform protein 
engineering experiments \cite{fersht}.}
\end{thebibliography}
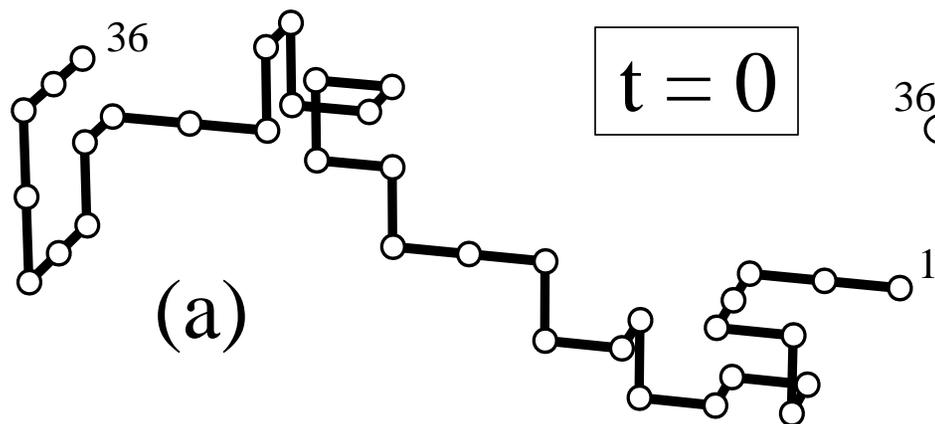
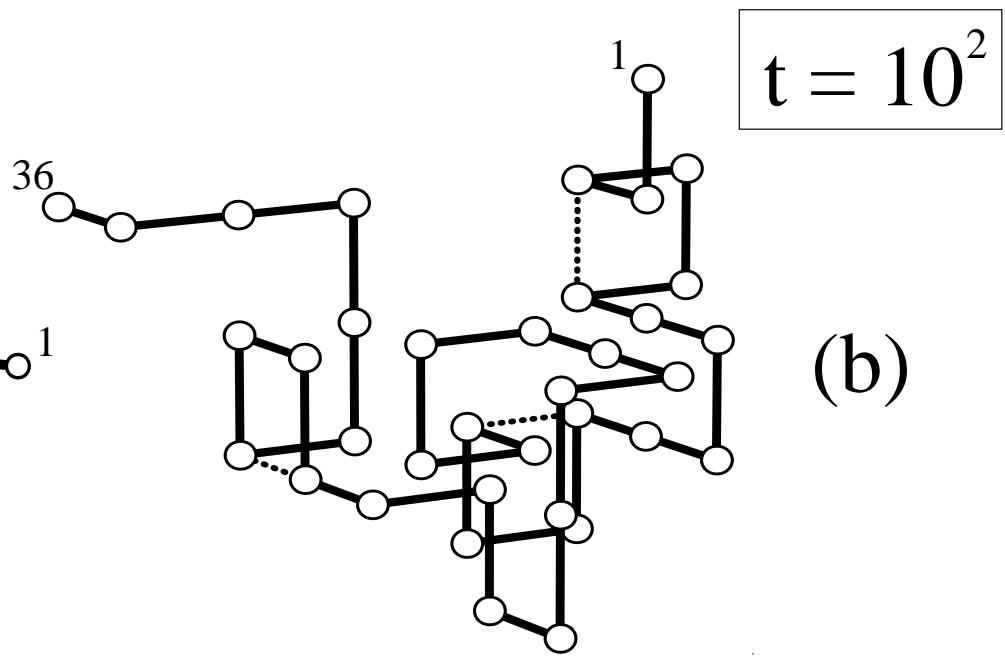
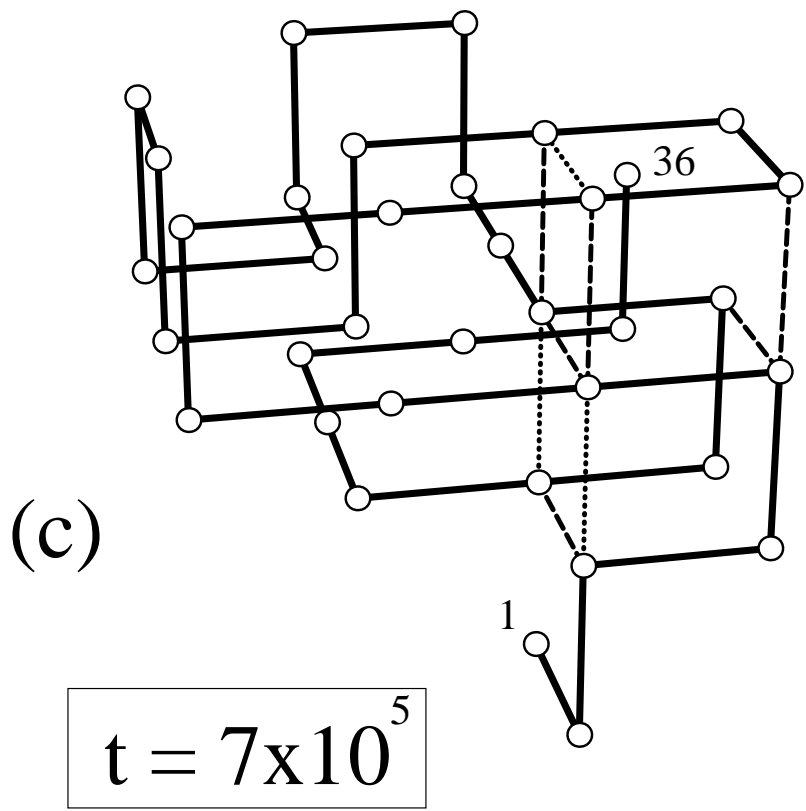
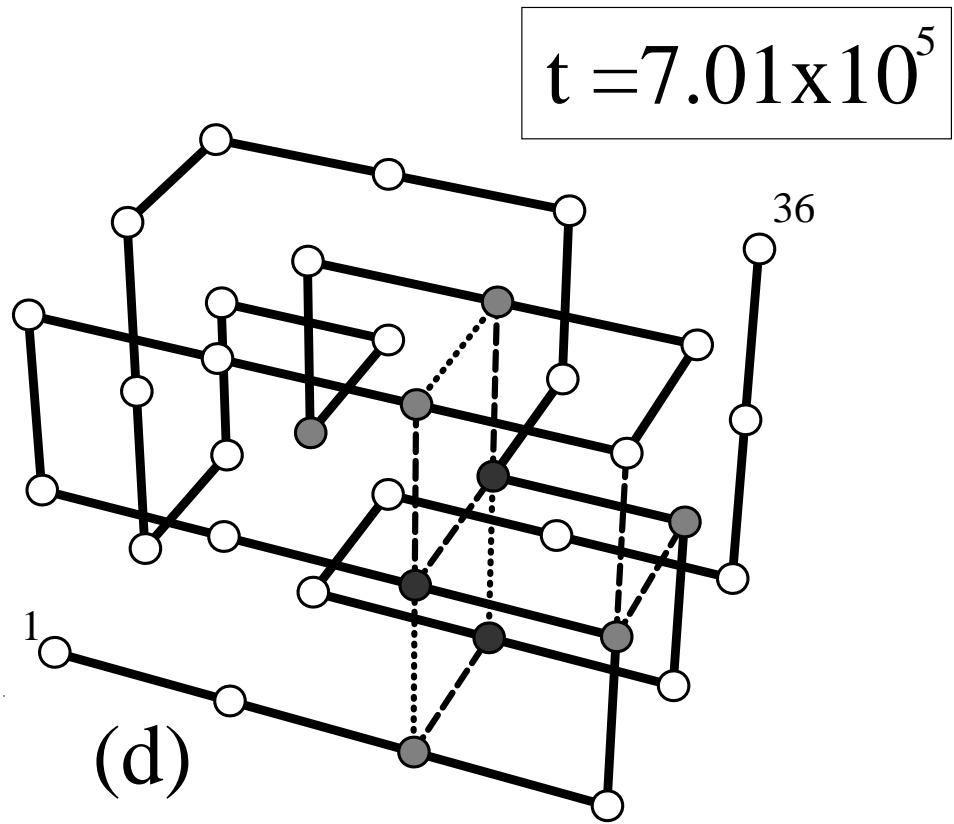
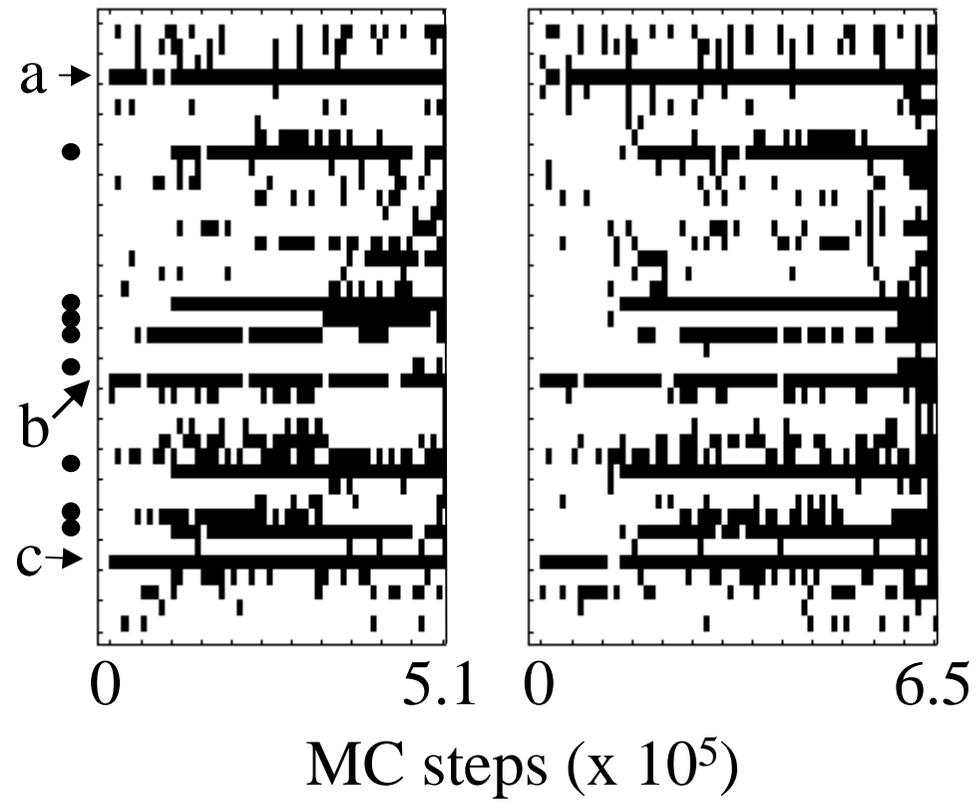
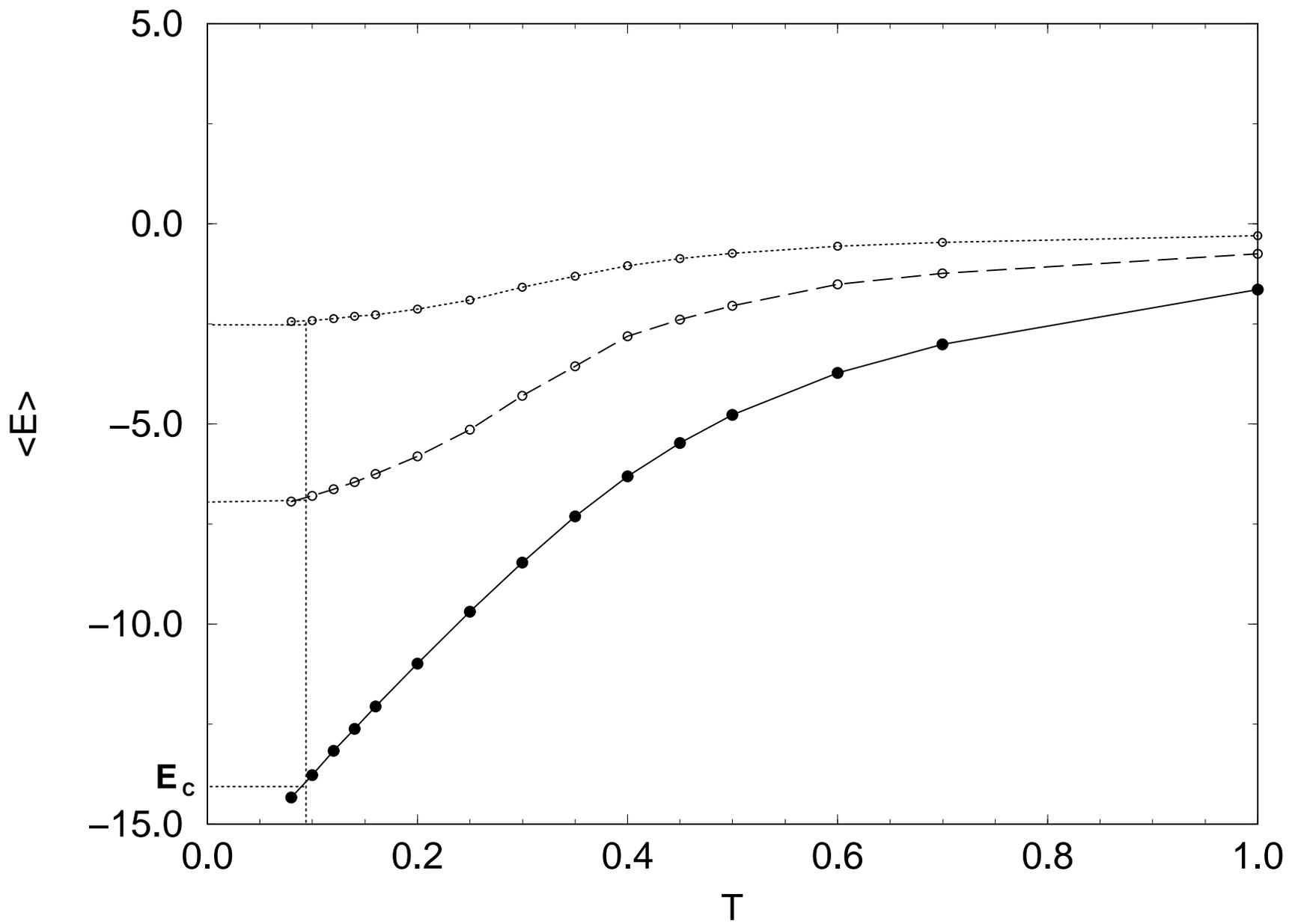
\end{document}